\documentstyle[a4,12pt,epsf]{article}

\newcommand{\be}{\begin{equation}}
\newcommand{\ee}{\end{equation}}
\newcommand{\bea}{\begin{eqnarray}}
\newcommand{\eea}{\end{eqnarray}}
\newcommand{\mb}{\overline{m}}

\begin{document}
\begin{titlepage}
\begin{flushright}
  KUNS-1918\\
  KAIST-TH-2004-06
\end{flushright}

\begin{center}
\vspace*{10mm}

{\LARGE \bf Quark masses and mixing angles in
heterotic orbifold models}
\vspace{12mm}

{\large
Pyungwon~Ko\footnote{E-mail address: pko@muon.kaist.ac.kr},
Tatsuo~Kobayashi\footnote{E-mail address:
  kobayash@gauge.scphys.kyoto-u.ac.jp}
~and~~Jae-hyeon~Park\footnote{E-mail address:
jhpark@muon.kaist.ac.kr}
}
\vspace{6mm}

{\it $^{1,3}$Department of Physics, KAIST, Daejon 305-701,
Korea}\\[1mm]

{\it $^2$Department of Physics, Kyoto University,
Kyoto 606-8502, Japan}

\vspace*{15mm}

\begin{abstract}
We study systematically the possibility for
realizing  realistic values of quark mass ratios
$m_c/m_t$ and $m_s/m_b$ and the mixing angle $V_{cb}$
by using only renormalizable Yukawa couplings derived from
heterotic orbifold models.
We assume one pair of up and down sector Higgs fields.
We find many realistic Yukawa matrices.

\end{abstract}

\end{center}
\end{titlepage}

\section{Introduction}

What is the origin of fermion masses and mixing angles is
one of important issues in particle physics.
They are determined by Yukawa couplings within the framework
of the standard model as well as its extension.
In a sense, $O(1)$ of Yukawa couplings seem natural. From
this viewpoint, how to derive hierarchically suppressed
Yukawa couplings is a key-point in understanding the
hierarchy of fermion masses and mixing angles.

Superstring theory is a promising candidate for unified theory.
Thus, it must predict fermion masses and mixing angles.
Actually, Yukawa couplings have been studied in several types of
4D string models,
that is, selection rules have been investigated and
$O(1)$ of Yukawa couplings have been calculated explicitly
in many 4D string models.
Among them, heterotic orbifold models \cite{Dixon:jw}
as well as intersecting
D-brane models are interesting,
because they lead to suppressed Yukawa couplings depending on
moduli \cite{Dixon:1986qv,Hamidi:1986vh,Ibanez:1986ka}.
Calculations of such moduli-dependent Yukawa couplings are
possible in orbifold models
\cite{Dixon:1986qv,Hamidi:1986vh,Burwick:1990tu,Kobayashi:2003vi},
since string theory can be solved on orbifolds.
Calculations of Yukawa couplings in intersecting
D-brane models are similar to those in heterotic orbifold
models \cite{Cremades:2003qj,Cvetic:2003ch,Abel:2003vv}.
Furthermore, the selection rule due to space group invariance
in orbifold models seems
unique \cite{Dixon:1986qv,Kobayashi:1991rp,Casas:1991ac},
e.g. compared with $Z_N$ discrete symmetries.
It allows non-trivially off-diagonal couplings.
Hence, orbifold models have a possibility for
leading to realistic mixing angles as well as fermion masses.
On the other hand, the selection rule for allowed couplings
in intersecting D-brane models is model-dependent.
As much as we are aware, there are no intersecting D-brane models
with realistic values of  mixing angles at tree-level with
the minimal number of Higgs fields.
Therefore, it is important to study systematically
the possibility for leading to realistic fermion masses
and mixing angles in heterotic orbifold models.

Indeed, a similar study has been done in Ref.~\cite{Casas:1992zt}.
However, the analysis in Ref.~\cite{Casas:1992zt}
concentrated rather only to
the second and third diagonal elements of Yukawa matrices,
that is, the mass ratios between the second and third families.
The other entries were assumed to be generated through
higher dimensional operators like the Froggatt-Nielsen
mechanism.
Indeed, it is plausible that nonrenormalizable couplings
play a role for suppressed entries in Yukawa matrices.
However, in realistic patterns of Yukawa matrices,
the (2,3) and (3,2) entries are the same as or larger than the
(2,2) entries.
Thus, if we assume that the (2,2) entries are originated from
stringy renormalizable couplings, it is natural to expect
the (2,3) and (3,2) entries are also obtained as
stringy renormalizable couplings.
In this paper we study the possibility for predicting
a realistic mixing angle as well as mass ratios.
We concentrate ourselves mainly to $(2 \times 2)$ sub-matrices
of the second and third quark families.
We study systematically the possibility for
obtaining realistic values of $V_{cb}$ and mass ratios
$m_c/m_t$ and $m_s/m_b$.
Then, we will show examples to lead to them.
To our knowledge, our result is the first examples, which
show explicitly the possibility for predicting
realistic values of mixing angles by use of only renormalizable
couplings in string models, when we consider
a pair of up and down Higgs fields, although
already there are proposals to introduce more Higgs fields
to lead to realistic Yukawa matrices \cite{Abel:2002ih}.

\section{Orbifold models and selection rule}

\subsection{Fixed points and twisted sectors}

Here we give a brief review on the structure of
fixed points on orbifolds and corresponding twisted states.
(See for their details Ref.~\cite{Kobayashi:1991rp,Casas:1991ac}.)
An orbifold is defined as a division of a torus
by a discrete rotation, i.e., a twist $\theta$.
For example, the 2D $Z_3$ orbifold is obtained by
dividing $R^2$ by the $SU(3)$ root lattice and its automorphism $\theta$,
that is, the Coxeter element of
$SU(3)$ algebra, which transforms the $SU(3)$ simple roots
$e_1$ and $e_2$,
\begin{equation}
\theta e_1 \rightarrow e_2,  \qquad \theta e_2 \rightarrow -e_1 -e_2.
\end{equation}
Thus, the twist  $\theta$ is the $Z_3$ rotation.
Similarly, the 2D $Z_6$ orbifold is obtained
as a division of $R^2$ by the $G_2$ lattice and
the $G_2$ Coxeter element, which transforms the $G_2$ simple roots
$e_1$ and $e_2$,
\begin{equation}
\theta e_1 \rightarrow -e_1-e_2,  \qquad \theta e_2 \rightarrow 3e_1 +2e_2,
\end{equation}
that is, the $Z_6$ twist.

The 6D $Z_3$ orbifold is a direct product of three 2D
$Z_3$ orbifolds.
Similarly, the so-called 6D  $Z_6$-I orbifold is
a direct product of two 2D $Z_6$ orbifolds and
one 2D $Z_3$ orbifold, that is,
eigenvalues of its twist $\theta$ are obtained as
$\theta = {\rm diag} (e^{2\pi i/6},e^{2\pi i/6},e^{-2\pi i/3})$.
Other 6D orbifolds are also defined in the same way
as a division of $R^6$ by a Lie lattice $\Lambda$ and its
Coxeter element $\theta$.\footnote{
See Ref.~\cite{Kobayashi:1991rp,Casas:1991ac} for details of Lie
lattices and their Coxeter elements, which
realize $Z_N$ orbifold twists.}

On an orbifold, there are two types of closed strings.
One type is untwisted strings, which are closed before
orbifold twisting, and
the other type is twisted strings.
The latter plays a important role here and
has the following twisted boundary condition,
\begin{equation}
X^i(\sigma =  2 \pi) = (\theta^k X)^i(\sigma = 0) + n e^i,
\end{equation}
where $e^i$ is the lattice vector defining $\Lambda$ and
$n$ is an integer.
Thus, ground states of $\theta^k$ twisted sectors are
assigned with fixed points $f$ on the orbifold, which are
defined as
\begin{equation}
f^i = (\theta^k f)^i + ne^i.
\end{equation}
This fixed point $f$ is presented by the corresponding
space group element $(\theta^k,ne^i)$.
For example, the 2D $Z_3$ orbifold has three fixed points
under $\theta$, and those are obtained as
\begin{equation}
g^{(0)}_{Z_3,1}=(0,0), \qquad
g^{(1)}_{Z_3,1}=(2/3,1/3),  \qquad
g^{(2)}_{Z_3,1}(1/3,2/3),
\end{equation}
in the $SU(3)$ simple root basis, up to
$(1 - \theta)\Lambda_{SU(3)}$.
Furthermore, those are denoted as
\begin{equation}
g^{(n)}_{Z_3,1}:~~(\theta,n e^1),
\end{equation}
where $n=0,1,2$.
The corresponding twisted grounds states are obtained as
$|g^{(n)}_{Z_3,1} \rangle$ with $n=0,1,2$.
Twenty seven fixed points on the 6D $Z_3$ orbifold are obtained
as direct products of three fixed points on the 2D orbifolds as
\begin{equation}
g^{(n)}_{Z_3,1} \otimes g^{(n')}_{Z_3,1}\otimes
g^{(n'')}_{Z_3, 1} ,
\end{equation}
where $n, n', n''=0,1,2$, and moreover the corresponding
twisted ground states are denoted as
\begin{equation}
| g^{(n)}_{Z_3,1} \rangle \otimes | g^{(n')}_{Z_3,1}
\rangle \otimes | g^{(n'')}_{Z_3,1} \rangle .
\end{equation}
$\theta^2$-twisted states are CPT conjugates of
$\theta$-twisted states.

Similarly, fixed points on the 2D $Z_6$ orbifold are obtained.
There is only one fixed point under $\theta$ on the 2D $Z_6$ orbifold,
that is,
\begin{equation}
g^{(0)}_{Z_6,1} = (0,0),
\end{equation}
in the $G_2$ simple root basis.
Furthermore, there are three fixed points under $\theta^2$,
\begin{equation}
g^{(0)}_{Z_6,2} = (0,0), \qquad g^{(1)}_{Z_6,2} = (0,1/3), \qquad
g^{(2)}_{Z_6,2} = (0,2/3) .
\end{equation}
Recall that these fixed points are defined up to the $G_2$ lattice.
\footnote{For example, for the fixed point $g^{(2)}_{Z_6,2} = (0,2/3)$, it
is often useful to use the following point,
$g^{(2)}_{Z_6,2} = (1,2/3) ,$
in order to calculate Yukawa couplings,
because this point is closer to the origin than $(0,2/3)$ and
we have exactly $\theta (0,1/3) = (1,2/3)$.}
The corresponding three twisted ground states are denoted as
$| g_{Z_6,2}^{(i)} \rangle$.
However, note that all of three points $g_{Z_6,2}^{(i)}$ are
not fixed under $\theta$.
While $g^{(0)}_{Z_6,2}$ is still a fixed point of $\theta$,
the two fixed points $g^{(1)}_{Z_6,2}$ and $g^{(2)}_{Z_6,2}$ are
transformed to each other by $\theta$.
Physical states consist of $\theta$-eigenstates.
Thus, we take linear combinations of states corresponding to
$g^{(1)}_{Z_6,2}$ and $g^{(2)}_{Z_6,2}$ as
\cite{Kobayashi:1990mc,Kobayashi:1991rp}
\begin{equation}
| g^{(1)}_{Z_6,2}; \pm 1 \rangle \equiv \frac{1}{\sqrt{2}}
\left(| g^{(1)}_{Z_6,2} \rangle
\pm | g^{(2)}_{Z_6,2} \rangle \right),
\label{lin-comb-2}
\end{equation}
with the eigenvalues $\gamma = \pm 1$, while
the state $| g^{(0)}_{Z_6,2} \rangle $ corresponding to
the fixed point $g^{(0)}_{Z_6,2}$ is a $\theta$-eigenstate.

In addition, there are four fixed points under $\theta^3$,
\begin{eqnarray}
& & g^{(0)}_{Z_6,3} = (0,0), \qquad g^{(1)}_{Z_6,3} = (0,1/2),
\nonumber \\
& & g^{(2)}_{Z_6,3} = (1/2,0), \qquad  g^{(3)}_{Z_6,3} = (1/2,1/2) .
\end{eqnarray}
Recall again that these fixed points are defined up to
the $G_2$ lattice.
\footnote{For Yukawa calculations, the following fixed points are often useful,
$g^{(0)}_{Z_6,3} = (0,0)$,
$g^{(1)}_{Z_6,3} = (1,1/2), $
$g^{(2)}_{Z_6,3} = (1/2,0),$
$g^{(3)}_{Z_6,3} = (1/2,1/2) $.
}
Not all of four points are fixed under $\theta$.
The $\theta$-eigenstates for each $G_2$ part are obtained as
\begin{equation}
 | g^{(0)}_{Z_6,3} \rangle, \qquad
 | g^{(1)}_{Z_6,3}; \gamma \rangle \equiv  \frac{1}{\sqrt{3}}
\left(| g^{(1)}_{Z_6,3} \rangle
+ \gamma  | g^{(2)}_{Z_6,3} \rangle + \gamma^2  | g^{(3)}_{Z_6,3}
 \rangle \right),
\end{equation}
where $\gamma = 1, \omega, \omega^2$ with $\omega = e^{2 \pi i/3}$.


Fixed points and twisted ground states for the 6D
$Z_6$-I orbifold are obtained as direct products of
two 2D $Z_6$ orbifolds and one 2D $Z_3$ orbifold.
The $\theta$ twisted sector has the following
fixed points,
\begin{equation}
g^{(0)}_{Z_6,1} \otimes g^{(0)}_{Z_6,1} \otimes g^{(i)}_{Z_3,1},
\end{equation}
for $i=0,1,2$, and the corresponding ground states are
denoted as
\begin{equation}
|g^{(0)}_{Z_6,1}\rangle \otimes |g^{(0)}_{Z_6,1} \rangle
\otimes | g^{(i)}_{Z_3,1} \rangle .
\end{equation}
The $\theta^2$ twisted sector has the following fixed points,
\begin{equation}
g^{(i)}_{Z_6,2} \otimes g^{(i')}_{Z_6,2} \otimes g^{(j)}_{Z_3,2},
\end{equation}
for $i,i',j=0,1,2$.
The $\theta$-eigenstates are obtained as
\begin{eqnarray}
& & | g^{(0)}_{Z_6,2}\rangle \otimes | g^{(0)}_{Z_6,2} \rangle
\otimes | g^{(j)}_{Z_3,2} \rangle,  \nonumber \\
& & | g^{(1)}_{Z_6,2};\gamma \rangle \otimes | g^{(0)}_{Z_6,2} \rangle
\otimes | g^{(j)}_{Z_3,2} \rangle,  \nonumber \\
& & | g^{(0)}_{Z_6,2}\rangle \otimes | g^{(1)}_{Z_6,2}; \gamma' \rangle
\otimes | g^{(j)}_{Z_3,2} \rangle,  \\
& & | g^{(1)}_{Z_6,2};\gamma\rangle \otimes | g^{(1)}_{Z_6,2}; \gamma' \rangle
\otimes | g^{(j)}_{Z_3,2} \rangle,  \nonumber
\end{eqnarray}
for $\gamma, \gamma' = \pm 1$.
The $\theta^3$ twisted sector has the following fixed points,
\begin{equation}
g^{(m)}_{Z_6,2} \otimes g^{(m')}_{Z_6,2},
\end{equation}
for $m,m'=0,1,2,3$.
The $\theta$-eigenstates are obtained as
\begin{eqnarray}
& & | g^{(0)}_{Z_6,3}\rangle \otimes | g^{(0)}_{Z_6,3} \rangle ,
\nonumber \\
& & | g^{(1)}_{Z_6,3};\gamma \rangle \otimes | g^{(0)}_{Z_6,3}
\rangle,
\nonumber  \\
& & | g^{(0)}_{Z_6,3}\rangle \otimes | g^{(1)}_{Z_6,3}; \gamma' \rangle,  \\
& & | g^{(1)}_{Z_6,3};\gamma\rangle \otimes | g^{(1)}_{Z_6,3};
\gamma' \rangle, \nonumber
\end{eqnarray}
where $\gamma, \gamma' = 1, \omega, \omega^2$.

Similarly, we can study fixed points and the structure of ground
states for other $Z_N$ orbifolds, and
those are explicitly shown in
Ref. \cite{Kobayashi:1991rp,Casas:1991ac}.
In what follows, the $\theta^k$ twisted sector of $Z_N$
orbifold models is denoted as $\hat T_k$.

\subsection{Selection rule}

Here we give a brief review on the selection rule for Yukawa couplings
in orbifold models.
(See for their details Ref.~\cite{Kobayashi:1991rp,Casas:1991ac}.)
The fixed point $f$ is denoted by its space group element,
$(\theta^k, (1-\theta^k) f)$, as said in the
previous subsection.
Thus, the three states corresponding to three fixed points
$(\theta^{k_i},(1-\theta^{k_i})f_i)$ for $i=1,2,3$ can couple if the
product of their space group elements
$\prod_i (\theta^{k_i},(1-\theta^{k_i})f_i)$  is
equivalent to identity.
Here, note that the fixed point $(\theta^k, (1-\theta^k) f)$
is equivalent to $(\theta^k, (1-\theta^k) (f+ \Lambda) )$,
that is, they belong to the same conjugacy class.
Thus, the Yukawa coupling among three states is allowed if
\begin{equation}
\prod_i (\theta^{k_i},(1-\theta^{k_i})f_i) =
(1,\sum_i (1-\theta^{k_i})\Lambda).
\end{equation}
This is called the space group selection rule \cite{Dixon:1986qv}.
That includes the point group selection rule, which
requires a product of twists, $\prod_i \theta^{k_i} $, to be identity.

For example, let us consider the $\hat T_1\hat T_1\hat T_1$ coupling
in the $Z_3$ orbifold models.
The $\hat T_1$ sector of the 2D orbifold has three fixed
points, $g^{(i)}_{Z3,1}$ $(i=0,1,2)$, and three states
corresponding fixed points $g^{(i_1)}_{Z3,1}$,
$g^{(i_2)}_{Z3,1}$ and $g^{(i_3)}_{Z3,1}$ can couple when
the following equation is satisfied,
\begin{equation}
i_1 + i_2 + i_3 = 0 ~~~({\rm mod~~~}3).
\end{equation}
Hence, couplings on the 2D $Z_3$ orbifold are diagonal, that is,
when we choose two states, the other state, which is allowed to
couple with them, is fixed uniquely.
Since the $\hat T_1$ sector on the 6D $Z_3$ orbifold is obtained
as a direct product of those on the 2D $Z_3$,
all of Yukawa couplings on the 6D $Z_3$ orbifold are
diagonal.
Therefore, we always obtain the following form of Yukawa matrix,
\begin{equation}
Y = \left(
\begin{array}{ccc}
Y_{11} & 0 & 0 \\
0 & Y_{22} & 0 \\
0 & 0 & Y_{33} \\
\end{array}
\right) ,
\label{diag-Y}
\end{equation}
when we consider only one pair of up and down
Higgs fields.
Thus, in this type of models we can not derive
non-vanishing mixing angles.
All of $Z_N$ orbifold models with $N=1$ 4D supersymmetry
are classified as $Z_3$, $Z_4$, $Z_6$-I, $Z_6$-II, $Z_7$,
$Z_8$-I, $Z_8$-II, $Z_{12}$-I and $Z_{12}$-II
orbifolds \cite{Dixon:jw,Ibanez:1987pj,Katsuki:1989bf}.
The situation in $Z_7$ orbifold models is the same as
one in $Z_3$ orbifold models,
and one can not derive non-vanishing mixing angles
only with renormalizable couplings and the
minimal number of up and down Higgs fields.

However, the situation is different for non-prime order orbifold
models, and off-diagonal couplings are allowed.
For example, let us consider $Z_6$-I orbifold models.
The 6D $Z_6$-I orbifold consists of $G_2 \times G_2 \times SU(3)$
part.
The $SU(3)$ part is the same as the $Z_3$ orbifold, and
only diagonal couplings are allowed.
Thus, this part is irrelevant to us, and
we concentrate to the $G_2 \times G_2$ part.
Among Yukawa couplings of twisted sectors,
the point group selection rule and $H$-momentum
conservation \cite{Friedan:1985ge}
allow only the following couplings \cite{Kobayashi:1990mc},
\begin{equation}
\hat T_1 \hat T_2 \hat T_3, \qquad \hat T_2 \hat T_2 \hat T_2 .
\end{equation}

For the $\hat T_1 \hat T_2 \hat T_3$ couplings, the space group
selection rule
requires only that the product of $\theta$-eigenvalues among
coupling states must be equal to identity, that is,
$\prod \gamma = 1$.
Thus, the twisted states relevant to allowed $\hat T_1 \hat T_2 \hat T_3$
couplings are the single $\hat T_1$ state,
\begin{equation}
| g^{(0)}_{Z_6,1} \rangle \otimes | g^{(0)}_{Z_6,1} \rangle ,
\end{equation}
and the five $\hat T_2$ states,
\begin{eqnarray}
\hat T_2^{(1)} &\equiv&
| g^{(0)}_{Z_6,2} \rangle \otimes | g^{(0)}_{Z_6,2} \rangle ,\nonumber \\
\hat T_2^{(2)} &\equiv&
| g^{(0)}_{Z_6,2} \rangle \otimes | g^{(1)}_{Z_6,2}; +1 \rangle ,
\nonumber \\
\hat T_2^{(3)} &\equiv&
| g^{(1)}_{Z_6,2}; +1 \rangle \otimes | g^{(0)}_{Z_6,2} \rangle ,
\label{T2-states} \\
\hat T_2^{(4,\gamma)} &\equiv&
| g^{(1)}_{Z_6,2}; \gamma \rangle \otimes | g^{(1)}_{Z_6,2};
\gamma^{-1} \rangle , \nonumber
\end{eqnarray}
where $\gamma = \pm 1$, and six $\hat T_3$ states,
\begin{eqnarray}
\hat T_3^{(1)} &\equiv&
| g^{(0)}_{Z_6,3} \rangle \otimes | g^{(0)}_{Z_6,3} \rangle ,
\nonumber \\
\hat T_3^{(2)} &\equiv&
| g^{(0)}_{Z_6,3} \rangle \otimes | g^{(1)}_{Z_6,3}; +1 \rangle ,
\nonumber \\
\hat T_3^{(3)} &\equiv&
| g^{(1)}_{Z_6,3}; +1 \rangle \otimes | g^{(0)}_{Z_6,3} \rangle ,
\label{T3-states} \\
\hat T_3^{(4,\gamma)} &\equiv&
| g^{(1)}_{Z_6,3}; \gamma \rangle \otimes | g^{(1)}_{Z_6,3};
\gamma^{-1} \rangle , \nonumber
\end{eqnarray}
where $\gamma = 1, \omega, \omega^2$.
Hence, in the case that fermions are assigned with $\hat T_2$ and $\hat T_3$
sectors and the Higgs is assigned with $\hat T_1$,
one can obtain non-trivial Yukawa matrices for three flavors,
whose determinant does not vanish and diagonalizing matrix
is not identity.

On the other hand, the space group selection for the
$\hat T_2\hat T_2\hat T_2$
couplings is exactly the same as one of $\hat T_1\hat T_1\hat T_1$ coupling
in $Z_3$ orbifold models, when we consider the basis of twisted states 
corresponding directly to fixed points.
However, in the $Z_6$-I orbifold models, we take linear combinations 
as Eq.~(\ref{lin-comb-2}), and such linear combinations can lead to 
non-trivial mixing.

We give a comment on the third plane, again.
In order to allow Yukawa couplings
through $\hat T_1 \hat T_2 \hat T_3$ couplings,
we have to assign the same fixed point on the $SU(3)$ plane
for  $\hat T_1$ and $ \hat T_2$, while the $SU(3)$ part is the fixed
torus for $\hat T_3$.
In this case, we just have $O(1)$ contribute, which are universal to
different flavors.
That implies that also for $\hat T_2 \hat T_2 \hat T_2$ couplings
the contribution from the third part is
universal, because we have to assign all of three families of
left-handed quarks to the same fixed point on the third plane.
That leads to an overall suppression factor or an overall factor of $O(1)$.
Anyway, that does not contribute to the
mixing angles or ratios of fermion masses.
We neglect this part,
and concentrate to the $G_2 \times G_2$
part for the $\hat T_2 \hat T_2 \hat T_2$ couplings, too.
Actually, in the following section
we assume all of relevant modes correspond to the same fixed point
on the $SU(3)$ plane.

Similarly, we can study Yukawa couplings for other non-prime order
$Z_N$ orbifold models.
In general, they allow off-diagonal couplings.
The numbers of twisted states relevant to allowed off-diagonal couplings
in $Z_4$, $Z_6$-II, $Z_8$-II $Z_{12}$-II orbifold models
are smaller than one in $Z_6$-I orbifold models.
In the next subsection,
we will mention
our reason why we do not study $Z_8$-I or $Z_{12}$-I orbifold models.
Thus, here we concentrate ourselves to analysis on
Yukawa matrices in $Z_6$-I models.
Yukawa matrices in other orbifold models will be studied
systematically elsewhere.

\subsection{Yukawa couplings}

The strength of Yukawa couplings has been calculated
by use of 2D conformal field theory.
It depends on locations of fixed points.
The Yukawa coupling strength of the
$\hat T_1 \hat T_2 \hat T_3$ coupling in $Z_6$-I
orbifold models is obtained for the $G_2 \times G_2 $ part 
as \cite{Dixon:1986qv,Hamidi:1986vh,Burwick:1990tu,Casas:1991ac}
\begin{equation}
Y = \sum_{f_{23}=f_2-f_3 +\Lambda} \exp [-\frac{\sqrt 3}{4 \pi} f_{23}^T
M f_{23}],
\label{Yukawa-123}
\end{equation}
up to an overall normalization factor, where
\begin{equation}
M = \left(
\begin{array}{cccc}
R_1^2 & -\frac{3}{2}R_1^2 & 0 & 0 \\ -\frac{3}{2}R_1^2 &
3R_1^2 & 0& 0 \\
0 & 0& R_2^2 & -\frac{3}{2}R_2^2 \\
0 & 0& -\frac{3}{2}R_2^2 & 3R_2^2
\end{array}
\right) ,
\label{M-matrix}
\end{equation}
in the $G_2 \times G_2$ root basis.
Here, $f_2$ and $f_3$ denote fixed points of
$\hat T_2$ and $\hat T_3$ sectors, respectively, and
$R_i$ corresponds to the radius of the $i$-th torus,
which can be written as a real part of the $i$-th
K\"ahler moduli $T_i$ up to a constant factor \footnote{
See Ref.~\cite{Lebedev:2001qg} for the proper normalization
of the moduli and Yukawa couplings
such that the transformation
$T_\ell \rightarrow T_\ell +i$ is a symmetry, that is, we
have the relation $Re(T_i)=\sqrt{3}R_i^2/(16 \pi^2)$.}, and
the imaginary part of $T_i$ can lead to CP phases.
However, we will concentrate ourselves to the $(2 \times 2)$
sub-matrices.
On the other hand, the full $(3\times 3)$ matrices are necessary
to study physical CP phase.
Thus, we do not consider imaginary parts of the K\"ahler moduli.\footnote{
Yukawa couplings also depend on
other moduli, e.g. continuous Wilson line moduli,
but here we consider only the dependence of $R_i$.}
The states with fixed points in the same conjugacy
class contribute to the Yukawa coupling.
Thus, we take summation of those contributions in eq.~(\ref{Yukawa-123}).
However, the states corresponding to the nearest
fixed points $(f_2,f_3)$ contribute dominantly
to the Yukawa coupling for a large value of $R_i$.
Hence, we calculate Yukawa couplings for the nearest
fixed points $(f_2,f_3)$.
We will give a comment on this point after showing examples
in the next section.

Similarly, the strength of $\hat T_2 \hat T_2 \hat T_2$
Yukawa couplings is obtained as
\begin{equation}
Y = \sum_{f_{23}= f_2-f_3 +\Lambda}
\exp [- \frac{\sqrt 3}{16 \pi}f_{23}^T M
f_{23}],
\label{Yukawa-222}
\end{equation}
where $M$ is the same matrix as Eq.(\ref{M-matrix}).
Here, $f_2$ and $f_3$ denote two of three fixed points
in $\hat T_2$ sectors.
Recall that when we choose two states, the other state,
which is allowed to couple, is uniquely fixed 
in the basis of states corresponding directly to 
fixed points.

Here, we give a comment on the K\"ahler metric.
It also depends on $R_i$, but its $R_i$ dependence
is the same for a $\hat T_k$ sector, while
that is different from another $\hat T_\ell$ sector $(k \neq \ell)$.
Thus, the $R_i$ dependence in the K\"ahler metric
is relevant only to the overall magnitude of Yukawa
matrices,\footnote{To be explicit, the $\hat T_1 \hat T_2 \hat T_3$
couplings have the factor $4Re(T_1) Re(T_2)(2Re(T_3))^{1/2}$ due to
the normalization of the K\"ahler metric, and
the $\hat T_2 \hat T_2 \hat T_2$
couplings have the factor $8 Re(T_1) Re(T_2) Re(T_3)$ due to
the normalization of the K\"ahler metric.
Furthermore, Yukawa couplings $\hat Y_{ijk}$
in global supersymmetric models
are related with Yukawa couplings $Y^{(SUGRA)}_{ijk}$ in supergravity as
$\hat Y_{ijk} = Y^{(SUGRA)}_{ijk}e^{\langle K \rangle /2}$
up to K\"ahler metric \cite{Soni:1983rm}.
Here,  $K$ is the K\"ahler potential and obtained as
$K = -\sum_i \ln (T_i + \bar T_i) + \cdots $,
where ellipsis denotes the contributions due to other fields with
large vacuum expectation values.
Hence, the K\"ahler potential $K$ itself is also relevant to
the overall magnitude of Yukawa matrices, but irrelevant to
mass ratios and mixing angles.}
but irrelevant to fermion mass ratios and mixing
angles when three families of quarks with the
same $SU(3)\times SU(2) \times U(1)_Y$ quantum numbers
are assigned with one of $\hat T_k$ sectors.
Therefore, only two parameters, $R_1$ and $R_2$,
are relevant to fermion mass ratios and mixing angles.

Similarly, we can calculate Yukawa couplings in generic
orbifold models, and those for all of $Z_N$ orbifold models
are shown in Ref.~\cite{Casas:1991ac}.
Off-diagonal couplings are originated from
the $SO(9)$ lattice part for the $Z_8$-I orbifold and
the $F_4$ lattice part for the $Z_{12}$-II orbifold.
Thus, in these models only one parameter, say $R_1^2$, corresponding
to the volume of the $SO(9)$ torus or the $F_4$ torus
is relevant to mass ratios and mixing angles,
while in $Z_6$-I models two parameters $R_1^2$ and $R_2^2$
are relevant.
That is the reason why we concentrate ourselves to
$Z_6$-I orbifold models here, but we will study
other $Z_N$ orbifold models elsewhere.

\section{Quark masses and mixing angles}

Here we study systematically the possibilities for
leading to realistic quark masses and the mixing angle
for the second and third families in $Z_6$-I
orbifold models by use of the structure of fixed points
and the strength of Yukawa couplings explained in the
previous section.
We assume the minimal number of up and down
Higgs fields.
Concerned about the other entries of Yukawa matrices,
we assume that Yukawa entries relevant to the
first family may be originated from higher
dimensional operators.

The experimental values are listed in Table~\ref{tab:nums},
and the ratios of running masses at the weak scale
are displayed in the last row of Table~\ref{tab:ex}.
\begin{table}[tb]
  \centering
  \begin{tabular}{c@{\hspace{2\tabcolsep}(}r@{.}l@{$\,\pm\,$}r@{.}l@{)}c|%
                  c@{\hspace{2\tabcolsep}(}r@{.}l@{$\,\pm\,$}r@{.}l@{)}c}
    \hline
    $\mb_u (m_W)$ &   1&7   & 0&5   & \ MeV &
    $\mb_d (m_W)$ &   3&7   & 0&9   & \ MeV \\
    $\mb_c (m_W)$ &   0&667 & 0&027 & \ GeV &
    $\mb_s (m_W)$ &   0&072 & 0&017 & \ GeV \\
    $\mb_t (m_W)$ & 175&3   & 5&1   & \ GeV &
    $\mb_b (m_W)$ &   2&906 & 0&047 & \ GeV \\
    \multicolumn{1}{ l}{$V_{cb}$} & \multicolumn{5}{r}{$(41.2\pm2.0)\times10^{-3}$} &
    \multicolumn{1}{|l}{$V_{us}$} & \multicolumn{5}{l}{$0.2196\pm0.0023$} \\
    \multicolumn{1}{ l}{$V_{ub}$} & \multicolumn{5}{r}{$(3.6\pm0.7)\times10^{-3}$} &
    \multicolumn{6}{|l}{} \\
    \hline
  \end{tabular}
  \caption{Input values used for fitting the moduli.
    The current quark masses in the $\overline{\rm MS}$ scheme
    at $m_W$ scale are
    from Ref.~\cite{Caravaglios:2002br}, and the CKM matrix elements
    from Ref.~\cite{Hagiwara:2002fs}. (See also Ref.~\cite{Fusaoka:1998vc}.)
    We supposed that $V_{ub}$ were real in the analysis.}
  \label{tab:nums}
\end{table}
The Yukawa couplings (\ref{Yukawa-123}),
(\ref{Yukawa-222}) are obtained at the Planck scale.
Precise values at low energy
depend on renormalization group flows, that is,
the matter content of models.
Thus, we do not try to derive exact values, but
we try to fit their orders.
We concentrate to the mass ratios, $m_c/m_t$ and $m_s/m_b$, and 
the mixing angle $V_{cb}$, but pay less attention to the 
overall magnitude like $Y_t$ and $Y_b$, because 
moduli values other than $R_1$ and $R_2$, in general, contribute 
to the overall magnitude as said in the previous section 
and we have ambiguity.

$Z_6$-I orbifold models have two types of Yukawa
couplings, $\hat T_1 \hat T_2 \hat T_3$ and
$\hat T_2 \hat T_2 \hat T_2$.
The $\hat T_1$ sector has a single relevant
state for the $G_2 \times G_2$ part, that is,
there is no variety for families.
That implies that we have to assign matter fields
with $\hat T_2$ and $\hat T_3$.
Hence, we have five classes of assignments,
which are shown in Table~\ref{tab:assign}.
Here, $Q$, $u$ and $d$ denote the left-handed
quarks, the up and down sector of right-handed
quarks, respectively.
In Assignments 1 and 2, both the up and down sectors
of Yukawa couplings are originated from
$\hat T_1 \hat T_2 \hat T_3$ couplings.
On the other hand, in Assignment 3, the up sector
Yukawa couplings are originated from
$\hat T_1 \hat T_2 \hat T_3$ couplings, and
the down sector Yukawa couplings are
originated from $\hat T_2 \hat T_2 \hat T_2$
couplings.
Oppositely, in Assignment 4, the up sector
Yukawa couplings are originated from
$\hat T_2 \hat T_2 \hat T_2$ couplings, and
the down sector Yukawa couplings are
originated from
$\hat T_1 \hat T_2 \hat T_3$ couplings.
In Assignment 5, both the up and down sectors of 
Yukawa couplings are originated from 
$\hat T_2 \hat T_2 \hat T_2$ couplings.

\begin{table}[tb]
  \centering
\begin{tabular}{|c|c|c|c|c|c|} \hline
Class & $Q$ & $u$ & $d$ & $H_u$ & $H_d$ \\ \hline \hline
Assignment 1 & $\hat T_2$ & $\hat T_3$ &  $\hat T_3$ &
 $\hat T_1$ & $\hat T_1$ \\ \hline
Assignment 2 & $\hat T_3$ & $\hat T_2$ &  $\hat T_2$ &
 $\hat T_1$ & $\hat T_1$ \\ \hline
Assignment 3 & $\hat T_2$ & $\hat T_3$ &  $\hat T_2$ &
 $\hat T_1$ & $\hat T_2$ \\ \hline
Assignment 4 & $\hat T_2$ & $\hat T_2$ &  $\hat T_3$ &
 $\hat T_2$ & $\hat T_1$ \\ \hline
Assignment 5 & $\hat T_2$ & $\hat T_2$ &  $\hat T_2$ &
 $\hat T_2$ & $\hat T_2$ \\ \hline
\end{tabular}
  \caption{Four classes of Assignments}
  \label{tab:assign}
\end{table}

Here we concentrate ourselves to the $\hat T_2$ and 
$\hat T_3$ states with $\gamma =1$.
The relevant number of $\hat T_2$ states
is 4 as Eq.(\ref{T2-states}). 
Thus, the possible number that
two flavors are assigned with $\hat T_2$ states
leading to different magnitudes of Yukawa couplings is
equal to 6.
Similarly, 
the possible number that
two flavors are assigned with $\hat T_3$ states
leading to different magnitudes of Yukawa couplings
is also  equal to 6.
Hence, in Assignment 1 there are
$6^3  = 216$ possibilities,
and Assignment 2 also has 216 possibilities.

For Assignment 3, there are  further four possibilities to 
assign $H_d$ with $\hat T_2$ states.
Totally, there are
$6^3 \times 4 = 864$.
Similarly, Assignment 4 has 864 possibilities.
Moreover, Assignment 5 has $6^3 \times 4^2 = 3456$.

We investigate systematically all of these
possibilities,\footnote{In practice, we have removed 
trivial possibilities, e.g. matrices with vanishing determinant and 
matrices with degenerate eigenvalues, before numerical study.}  
varying two
independent parameters $R_1$ and $R_2$.
We find many configurations leading
to realistic values of $m_c/m_t$, $m_s/m_b$ and
$V_{cb}$, which are consistent with experimental values 
up to $O(1)$ factor.
In particular, the numbers of realistic examples in 
Assignments 1 and 2 are larger than those in other assignments.
Here we show one of the best fitting examples in each class of 
Assignment.
Table~\ref{tab:ex} shows examples leading to realistic values of
$m_c/m_t$, $m_s/m_b$ and $V_{cb}$.
The first column shows the class of Assignments.
The second column shows assignments of quarks and Higgs fields
with twisted states.
The third and fourth columns show the values of
$R_1^2$ and $R_2^2$ corresponding to the best fit with
the experimental values.
The last three columns show predicted values of
$m_c/m_t$, $m_s/m_b$ and $V_{cb}$.
We have also studied the cases including $\hat T_2$ states with 
$\gamma \neq 1$, but we have not obtained more realistic results 
than those shown in Table~\ref{tab:ex}.
We do not need to consider $\hat T_3$ states with $\gamma \neq 1$,
since they lead to the same strength of $\hat T_1 \hat T_2 \hat T_3$
type Yukawa coupling as $\gamma = 1$.

The second row in Table~\ref{tab:ex} shows an example
in the class of Assignment 1.
Its explicit form of up and down Yukawa matrices are
obtained as
\begin{equation}
  \label{eq:yukawa-a1-72}
  Y_u = \left(\begin{array}{ll}
      0.0416 & 0.718 \\
      0.0557 & 0.848
    \end{array}\right) , \qquad
  Y_d = \left(\begin{array}{ll}
      0.0313 & 0.0416 \\
      0.0370 & 0.0557
  \end{array}\right) .
\end{equation}
We have neglected a common factor of $O(1)$.
In this example, the down sector corresponds to a democratic form, 
while the up sector is hierarchical.
We have the ratio $Y_t/Y_b = 13$.
In the class of Assignment 1, there are many examples
leading to similarly realistic results for $m_c/m_t$, $m_s/m_b$ and $V_{cb}$,
with both $Y_t/Y_b =O(1)$ and 
$Y_t/Y_b =O(10)$, including the case with different 
overall magnitude.

The third row in Table~\ref{tab:ex} shows an example
in the class of Assignment 2.
The explicit Yukawa matrices are
obtained as
\begin{equation}
  \label{eq:yukawa-a2-76}
  Y_u = \left(\begin{array}{ll}
      0.0281 & 0.439 \\
      0.0371 & 0.665
    \end{array}\right) , \qquad
  Y_d = \left(\begin{array}{ll}
      0.0199 & 0.0281 \\
      0.0302 & 0.0371
  \end{array}\right) .
\end{equation}
This form is similar to Eq.~(\ref{eq:yukawa-a1-72})
and leads to the ratio $Y_t/Y_b = 14$.
In the class of Assignment 2, we have many examples
leading to similarly realistic values of $m_c/m_t$, $m_s/m_b$ and $V_{cb}$,
with both $Y_t/Y_b =O(1)$ and 
$Y_t/Y_b =O(10)$, including the case with different 
overall magnitude.


The fourth row in Table~\ref{tab:ex} shows an example in
the class of Assignment 3.
The explicit Yukawa matrices are obtained as
\begin{equation}
  \label{eq:yukawa-a3-52}
  Y_u = \left(\begin{array}{ll}
      0.0000636 & 2.61 \times 10^{-7} \\
      0.0000344 & 0.0168
    \end{array}\right) , \qquad
  Y_d = \left(\begin{array}{ll}
      0 & 0.0251 \\
      0.225 & 0.500
  \end{array}\right) .
\label{assign-31}
\end{equation}
This leads to realistic values for $m_c/m_t$, $m_s/m_b$ and
$V_{cb}$, but the ratio $Y_t/Y_b$ is small.
If we change the configuration on the $SU(3)$ part for the
down sector such that $Q$, $d$ and $H_d$ correspond to different 
fixed points on the $SU(3)$ part, 
we can obtain a small value of $Y_b$.
However, this example leads to too much suppressed top Yukawa coupling.
Similarly, in Assignment 3, most of examples leading to realistic values of 
$m_c/m_t$, $m_s/m_b$ and $V_{cb}$ predict the top Yukawa coupling, 
which is smaller than $O(1)$.
Thus, these examples  may lead to smaller top mass than
the experimental value.
One solution for this problem is to enhance
the overall magnitude of Yukawa couplings
by choosing suitable values of moduli, which
contribute only the overall size of Yukawa couplings
(through the K\"ahler potential),
but not ratios or mixing angles.


The fifth row in Table~\ref{tab:ex} shows an example in
the class of Assignment 4.
The explicit form of Yukawa matrices in this example are
obtained as
\begin{equation}
 Y_u =  \left(
 \begin{array}{cc}
 0 & 0.00569 \\
 0.0179  & 0.159
 \end{array}
 \right), \qquad
 Y_d =  \left(
 \begin{array}{cc}
 0.000133 & 3.83\times 10^{-7} \\
 2.88 \times 10^{-12} & 0.00379
 \end{array}
 \right) .
 \end{equation}
This leads to the realistic mass ratios and mixing able, but 
the small overall magnitude similarly to the 
example in Assignment 3.
We need some enhancement of the overall magnitude.

The sixth row in Table~\ref{tab:ex} shows an example in
the class of Assignment 5.
The explicit form of Yukawa matrices in this example are
obtained as
\begin{equation}
 Y_u =  \left(
 \begin{array}{cc}
 0 & 0.0309 \\
 0.0309  & 0.500
 \end{array}
 \right), \qquad
 Y_d =  \left(
 \begin{array}{cc}
 0.00132 & 0 \\
 0.0214 & 0.0309
 \end{array}
 \right) .
 \end{equation}
This leads to the ratio $Y_t/Y_b=13$.
In Assignment 5, we also have examples leading to 
similarly realistic results for $m_c/m_t$, $m_s/m_b$ and 
$V_{cb}$, and the ratio $Y_t/Y_b=O(1)$.

\begin{table}[tb]
  \centering
\small
\begin{tabular}{|c|ccccc|c|c|c|c|c|} \hline
Class & $Q_2,Q_3$ & $u_2,u_3$ & $d_2,d_3$ & $H_u$ & $H_d$ &
$(R_1)^2$ & $(R_2)^2$ & $\frac{m_c}{m_t}$ &  $\frac{m_s}{m_b}$ &
$V_{cb}$  \\ \hline \hline
1 & $\hat T_2^{(2)}, \hat T_2^{(4)}$ &
$\hat T_3^{(3)}, \hat T_3^{(2)}$ &
$\hat T_3^{(1)}, \hat T_3^{(3)}$ &
 $\hat T_1$ & $\hat T_1$ &
27.8 & 107 & 0.0038 & 0.029 & 0.041 \\ \hline
2 & $\hat T_3^{(2)}, \hat T_3^{(4)}$ &
$\hat T_2^{(3)}, \hat T_2^{(2)}$ &
$\hat T_2^{(1)}, \hat T_2^{(3)}$ &
 $\hat T_1$ & $\hat T_1$ &
24.0 & 150 & 0.0038 & 0.032 & 0.041 \\ \hline
3 & $\hat T_2^{(1)}, \hat T_2^{(4)}$ &
$\hat T_3^{(2)}, \hat T_3^{(4)}$ &
$\hat T_2^{(2)}, T_2^{(4)}$ &
 $\hat T_1$ & $\hat T_2^{(4)}$ &
196 & 316 & 0.0038 & 0.019 & 0.042 \\ \hline
4 & $\hat T_2^{(2)}, \hat T_2^{(4)}$ &
$\hat T_2^{(2)}, \hat T_2^{(3)}$ &
$\hat T_3^{(1)}, \hat T_3^{(4)}$ &
 $\hat T_2^{(4)}$ & $\hat T_1$ &
416 & 226 & 0.0040 & 0.035 & 0.035
\\ \hline 
5 & $\hat T_2^{(2)}, \hat T_2^{(4)}$ &
$\hat T_2^{(2)}, \hat T_2^{(4)}$ &
$\hat T_2^{(3)}, \hat T_2^{(2)}$ &
 $\hat T_2^{(4)}$ & $\hat T_2^{(4)}$ &
368 & 400 & 0.0038 & 0.029 & 0.041
\\ \hline \hline\multicolumn{8}{|c|}{Central values from experiments} & 0.0038 & 0.025 & 0.041
\\
\hline
\end{tabular}
  \caption{Realistic examples}
  \label{tab:ex}
\end{table}

As results, we have found many examples of assignments
leading to realistic values of the mass ratios $m_c/m_t$ and $m_s/m_b$ and
the mixing angle $V_{cb}$.
It is quite non-trivial to derive reasonable values of
three observables $m_c/m_t$, $m_s/m_b$ and $V_{cb}$ by only two
independent parameters $R_1$ and $R_2$ in models
with renormalizable couplings, which can be derived from
string models.

So far, we have considered the Yukawa couplings which
are induced only for the nearest fixed points.
Such approximation is reliable in our results, because
our realistic examples are obtained for large values of
$R_i$.
Actually, we have examined Yukawa coupling contributions
due to quite far fixed points.
Then we have obtained almost same results.

We can extend the above analysis to the full $(3 \times 3)$
Yukawa matrices.
For example, we assign three families of $Q$, $u$, $d$ as follows,
\begin{eqnarray}
Q &:& \hat T_3^{(1)}, T_3^{(2)}, T_3^{(4)}, \nonumber \\
u &:& \hat T_2^{(1)}, T_2^{(2)}, T_2^{(3)}, \\
d &:& \hat T_2^{(2)}, T_2^{(3)}, T_2^{(4)},  \nonumber
\end{eqnarray}
and both of Higgs fields with $\hat T_1$.
Then, this example leads to the following mass ratios and
mixing angles,
\begin{eqnarray}
\frac{m_u}{m_t}&=& 3.7 \times 10^{-9} ,
\qquad \frac{m_c}{m_t}=3.8 \times 10^{-3}, \nonumber \\
\frac{m_d}{m_b}&=& 1.7 \times 10^{-3} ,
\qquad \frac{m_s}{m_b}=9.9 \times 10^{-3},\\
V_{us} &=& 0.22, \qquad V_{cb} = 4.6 \times 10^{-9},
\qquad V_{ub} = 4.8 \times 10^{-9}, \nonumber
\end{eqnarray}
when we take $R_1^2=37.1$ and $R_2^2=572$.
In this example,
all of mass ratios except $m_u/m_t$ are reasonable values,
and the Cibibbo angle $V_{us}$ is predicted as a realistic value
by only two parameters, $R_1$ and $R_2$,
but the other mixing angles are suppressed too much,
although it is non-trivial to fix seven observables by only two parameters.
Similarly, we have investigated all of possibilities,
but it seems difficult to fit all of quark masses and mixing
angles to be consistent with the experimental values
by only two parameters $R_1$ and $R_2$.
Some extension is necessary in $Z_6$-I orbifold models in order to
derive all of quark masses and mixing angles.
It may be reasonable that higher dimensional operators
can contribute to small couplings \footnote {See for related subjects
e.g. Ref.~\cite{nonr}.} like
Yukawa matrix entries relevant to the first family, as we assumed
in the analysis for $(2 \times 2)$ sub-matrices.
It is also plausible that loop-effects have non-trivial
contributions for some entries.
Alternatively, it would be helpful to introduce more than one pair of
$H_u$ and $H_d$.

\section{Conclusion}

We have systematically studied the possibility for
leading to realistic values of $m_c/m_t$, $m_s/m_b$ and
$V_{cb}$ in $Z_6$-I orbifold models.
We have found realistic examples of Yukawa matrices.
In particular, the classes of Assignments 1 and 2 have 
many realistic Yukawa matrices.
Our result is the first examples to show the possibility
for deriving the realistic mixing angle by renormalizable
couplings in string models with one pair of $H_u$ and $H_d$.

To realize our results, the moduli  $R_1$ and $R_2$
must be stabilized at proper values.
How to stabilize these moduli is an important issue to
study further.

One can extend our analysis to other non-prime order $Z_N$
orbifold models.
Similarly we can discuss $Z_N \times Z_M$ orbifold models,
and extensions to non-supersymmetric orbifold models might also
be interesting.
Such systematical study will be done elsewhere.

Another important extension is to study the lepton sector.
The situation would change for realizing large mixing angles.
It is interesting to investigate systematically whether
one can obtain realistic lepton masses and mixing angles
by renormalizable couplings derived from string models.
Such systematical analysis will also be done elsewhere.

\section*{Acknowledgment}

The authors would like to thank Oleg Lebedev for useful discussions.
One of the authors (T.~K.) would like to
thank hospitality of KAIST, where
a part of this work was studied.
T.~K.\/ is supported in part by the Grant-in-Aid for
Scientific Research  (\#16028211) and
the Grant-in-Aid for
the 21st Century COE ``The Center for Diversity and
Universality in Physics'' from the Ministry of Education, Culture,
Sports, Science and Technology of Japan.
PK and JP are supported in part by
KOSEF Sundo Grant R02-2003-000-10085-0,
KRF grant KRF-2002-070-C00022,
BK21 Haeksim program and
KOSEF SRC program through CHEP at Kyungpook National University.

\end{document}